\documentclass[aps,showpacs,preprintnumbers,amsmath, amssymb]{revtex4}

\usepackage{float}
\usepackage{graphics,epsfig}
\usepackage{graphicx}
\usepackage{dcolumn}
\usepackage{bm}

\begin{document}
\baselineskip=0.8 cm

\title{{\bf Hair mass bound in the black hole with non-zero cosmological constants}}
\author{Yan Peng$^{1}$\footnote{yanpengphy@163.com}}
\affiliation{\\$^{1}$ School of Mathematical Sciences, Qufu Normal University, Qufu, Shandong 273165, China}

\vspace*{0.2cm}
\begin{abstract}
\baselineskip=0.6 cm
\begin{center}
{\bf Abstract}
\end{center}

We study mass bounds of Maxwell fields in RN black holes and genuine hair
in Einstein-Born-Infeld black holes with various cosmological constants.
It shows that the Maxwell field serves as a good probe to disclose the
hair distribution described with the event horizon and
the photonsphere. And we find that the Hod's lower bound
obtained in asymptotically flat space also holds in the
asymptotically dS Einstein-Born-Infeld hairy black holes.
In contrast, the Hod's lower bound can be invaded in the
asymptotically AdS Einstein-Born-Infeld hairy black holes.
It implies that the AdS boundary could make the Born-Infeld hair
easier to condense in the near horizon area.

\end{abstract}

\maketitle
\newpage
\vspace*{0.2cm}

\section{Introduction}

The famous no hair conjecture of Wheeler \cite{RR,JDB,PW}
was motivated by researches on uniqueness theorems that
Einstein-Maxwell black holes are
described by only three conserved parameters:
mass, electric charge and angular momentum \cite{WI1,BC,SW,DC2,JI}.
The belief in the no hair conjecture was based on a simple
physical picture that matter fields outside black holes would eventually
be radiated away to infinity or be swallowed by
the black hole horizon except when those fields were associated
with the three conserved parameters.
In accordance with this logic, stationary black holes indeed
exclude the existence of scalar fields, massive
vector fields and spinor fields in the exterior
spacetime of black holes \cite{JE1,JE2,JE3,BM2,BM3,JH1,JH2}.

However, nowadays we are faced with the surprising discovery of various types of
hairy black holes. The first of which were Einstein-Yang-Mills black holes \cite{PBH1,PBH2,PBH3}.
After that, other static hairy black hole solutions were also discovered
in theories like Einstein-Skyrme, Einstein-non-Abelian-Proca, Einstein-Yang-Mills-Higgs
and Einstein-Yang-Mills-Dilaton and hair formation in non-static kerr black holes was investigated,
for references see \cite{H1}-\cite{H18} and reviews \cite{H19,H20}.
The discovery of front hairy black holes provides a challenge
to the validity of the classical no hair theorem.
Now, it is clear that the formation of hair is due to the fact that self-interaction can
bind together the hair in a region very close to the horizon and another region relatively distant
from the horizon \cite{DNH}.
In accordance with this physical picture, a no short hair
theorem was proposed as an alternative to the no hair conjecture based on the fact that
the black hole hair of Einstein-Yang-Mills
fields must extend above the photonsphere \cite{DNH}.
Shahar Hod also proved a no short scalar hair theorem that
linearized massive scalar fields have no short hair behaviors
in non-spherically symmetric non-static kerr black holes \cite{SHA}.

Along this line, it is interesting to study the distribution of hair mass.
For the limit case of the linear Maxwell field,
Hod showed that the region above the photonsphere
contains at least half of the total mass of Maxwell fields
and also found that this lower bound
holds for various genuine hairy black holes
in Einstein-Yang-Mills, Einstein-Skyrme, Einstein-non Abelian-Proca,
Einstein-Yang-Mills-Higgs and Einstein-Yang-Mills-Dilaton systems \cite{SH}.
And it was found that the non-linear Einstein-Born-Infeld black holes also satisfy
this lower bound that half of the Born-Infeld hair is above the photonsphere \cite{YSM}.
In fact, it is reasonable to use Maxwell fields to study density distribution
of genuine hair since the Maxwell field case is a linear
limit of the non-linear Einstein-Born-Infeld theory.
The front studies of hair mass bounds were carried out in asymptotically flat backgrounds.
As a further step, it is meaningful to extend the discussion in asymptotically flat black holes
to spacetimes with non-zero cosmological constants.

In the following, we introduce black holes with non-zero cosmological constants
and obtain bounds for linear hair mass ratio.
We also disclose properties of genuine hair in Einstein-Born-Infeld black holes.
We will summarize our main results at the last section.

\section{Analytical studies of linear hair mass bounds}

In this paper, we use the Maxwell field as a linear limit to disclose
properties of genuine hair similar to approaches in \cite{SH}.
And the four dimensional Einstein-Maxwell black hole geometries with
non-zero cosmological constant $\Lambda$ are described by \cite{ZZ1,ZZ2,ZZ3,EW}:
\begin{eqnarray}\label{AdSBH}
ds^{2}&=&-f(r)dt^{2}+f(r)^{-1}dr^{2}+r^{2}(d\theta^2+sin^{2}\theta d\phi^{2}),
\end{eqnarray}
where metric functions $f(r)=1-\frac{2M}{r}+\frac{Q^2}{r^2}-\Lambda r^2$
with M as the ADM mass and Q as the charge.

The mass $m(r)$ of the Maxwell field above the radius r is given by
\begin{eqnarray}\label{AdSBH}
m(r)=\int_{r}^{+\infty}4\pi r'^{2}\rho(r')dr'.
\end{eqnarray}
For the Maxwell field, one has the energy density $\rho(r)=-T_{t}^{t}=\frac{Q^2}{8\pi r^4}$ \cite{SH}.
It yields $m(r)=\frac{Q^2}{2r}$ for the mass function.

It was found that the photonsphere can be conveniently used to
describe spatial distribution of the matter field \cite{DNH,SH}.
According to the approach in \cite{SH}, the radius $r_{\gamma}$
of the null circular geodesic (photonsphere) in the RN black
hole is determined by the relation
\begin{eqnarray}\label{AdSBH}
2f(r_{\gamma})-r_{\gamma}f'(r_{\gamma})=0.
\end{eqnarray}
From (3), one obtains the radius $r_{\gamma}$ independent of the cosmological constants as
\begin{eqnarray}\label{AdSBH}
r_{\gamma}=\frac{1}{2}(3M+\sqrt{9M^2-8Q^2}).
\end{eqnarray}

We define $r_{H}$ as the black hole event horizon satisfying $f(r_{H})=0$.
And an interesting quantity which characterizes the spatial
distribution of the hair is given by the dimensionless hair mass ratio
$\frac{m^{+}_{hair}}{m^{-}_{hair}}$, where
\begin{eqnarray}\label{AdSBH}
m^{+}_{hair}=m(r_{\gamma})
\end{eqnarray}
is the mass of the hair above the photonsphere
and
\begin{eqnarray}\label{AdSBH}
m^{-}_{hair}=m(r_{H})-m(r_{\gamma})
\end{eqnarray}
is the mass of the hair contained between the
event horizon and the photonsphere.
For the linear hair of Maxwell field outside
the Reissner-Nordstr$\ddot{o}$m (RN) black hole,
Hod obtained bounds
on the ratio $\frac{m^{+}_{hair}}{m^{-}_{hair}}\geqslant 1$.
The ratio can be expressed as
\begin{eqnarray}\label{AdSBH}
\frac{m^{+}_{hair}}{m^{-}_{hair}}=\frac{\frac{Q^2}{2r_{\gamma}}}{\frac{Q^2}{2r_{H}}-\frac{Q^2}{2r_{\gamma}}}
=\frac{1}{\frac{r_{\gamma}}{r_{H}}-1}.
\end{eqnarray}

We have mentioned that $r_{\gamma}=\frac{1}{2}(3M+\sqrt{9M^2-8Q^2})$ is independent of the cosmological
constant $\Lambda$. In order to study the ratio $\frac{m^{+}_{hair}}{m^{-}_{hair}}$,
we try to research on how the cosmological constant can affect the event horizon $r_{H}$.
The event horizon $r_{H}$ can be obtained from the equation
$1-\frac{2M}{r}+\frac{Q^2}{r^2}-\Lambda r^2=\frac{1}{r^2}(r^2-2Mr+Q^2-\Lambda r^4)=0$
or the equation
\begin{eqnarray}\label{AdSBH}
r^2-2Mr+Q^2-\Lambda r^4=0.
\end{eqnarray}

Caes I: $\Lambda>0$

For the case of $\Lambda>0$ or asymptotically dS black hole spacetime,
the equation (8) has three real positive roots $r_{h}$, $r_{H}$ and $r_{0}$,
where $r_{h}$ is the Cauchy horizon, $r_{H}$ is the event
horizon and $r_{0}$ is the cosmological horizon with $r_{h}<r_{H}<r_{0}$ \cite{ZZ1,ZZ2}.
For simplicity, we introduce a function $y=r^2-2Mr+Q^2-\Lambda r^4$.
And the event horizon $r=r_{H}$ can be obtained from $y=0$.
In this work, we are interested in the case of $M\geqslant Q$.
According to the relation $r_{H}^2-2M r_{H}+Q^2=\Lambda r_{H}^4>0$,
there is $r_{H}>M+\sqrt{M^2-Q^2}$ or $r_{H}<M-\sqrt{M^2-Q^2}$.
Since $y'=2(r-M)-4 \Lambda r^3<0$ for $r\leqslant M$,
at most one root of $y=0$ is in the range of $r\leqslant M$
and there is $r_{H}>M$. Then we further have $r_{H}>M+\sqrt{M^2-Q^2}$.

Considering the fact $r_{H}>M+\sqrt{M^2-Q^2}$, the mass ratio can be expressed with $x=\frac{Q}{M}\in[0,1]$ as
\begin{eqnarray}\label{AdSBH}
\frac{m^{+}_{hair}}{m^{-}_{hair}}=\frac{1}{\frac{r_{\gamma}}{r_{H}}-1}>\frac{1}{\frac{\frac{1}{2}(3M+\sqrt{9M^2-8Q^2})}{M+\sqrt{M^2-Q^2}}-1}
=\frac{1}{\frac{\frac{1}{2}(3+\sqrt{9-8x^2})}{1+\sqrt{1-x^2}}-1}.
\end{eqnarray}

According to the fact that
\begin{eqnarray}\label{AdSBH}
(\frac{1}{\frac{\frac{1}{2}(3+\sqrt{9-8x^2})}{1+\sqrt{1-x^2}}-1})'_{x}=-\frac{2x}{(\frac{(\frac{1}{2}(3+\sqrt{9-8x^2})}{1+\sqrt{1-x^2}}-1)^{2}}
\frac{17-18x^2+3\sqrt{9-8x^2}+8\sqrt{1-x^2}}{3\sqrt{9-8x^2}+8\sqrt{1-x^2}}
\end{eqnarray}
and
\begin{eqnarray}\label{AdSBH}
17-18x^2+3\sqrt{9-8x^2}+8\sqrt{1-x^2}\geqslant 17-18+3+0=2>0,
\end{eqnarray}
we have $(\frac{1}{\frac{\frac{1}{2}(3+\sqrt{9-8x^2})}{1+\sqrt{1-x^2}}-1})'_{x}<0$ for all $x\in[0,1]$.
So we have
\begin{eqnarray}\label{AdSBH}
\frac{m^{+}_{hair}}{m^{-}_{hair}}> 1
\end{eqnarray}
and the lower bound is with $x=1$ and $\Lambda\rightarrow 0$.

From the relation $r_{H}^2-2M r_{H}+Q^2-\Lambda r_{H}^4=0$,
we arrive at $\Lambda=\frac{r_{H}^2-2M r_{H}+Q^2}{r_{H}^4}$.
Then there is $\frac{d\Lambda}{dr_{H}}=\frac{4(r_{H}^2-2M r_{H}+Q^2)}{r_{H}^5}
+\frac{2r_{H}-2M}{r_{H}^4}> 0$ since we have proved $r_{H}>M+\sqrt{M^2-Q^2}$.
So $\Lambda$ is an increasing function of $r_{H}$
and the event horizon $r_{H}$ also increases when we increase the value of $\Lambda$.
For $r_{H}=r_{\gamma}$ or $\Lambda=\frac{r_{\gamma}^{2}-2M r_{\gamma}+Q^2}{r_{\gamma}^{4}}>0$
with $r_{\gamma}=\frac{1}{2}(3M+\sqrt{9M^2-8Q^2})$, we have
\begin{eqnarray}\label{AdSBH}
\frac{m^{+}_{hair}}{m^{-}_{hair}}=\frac{1}{\frac{r_{\gamma}}{r_{H}}-1}=\frac{1}{\frac{r_{H}}{r_{H}}-1}=+\infty.
\end{eqnarray}

In all, the linear hair mass ratio satisfies the Hod's mass bound.
We further conjecture that asymptotically
dS genuine hairy black holes may also obey the
Hod's hair mass bound.

Caes II: $\Lambda<0$

For another case of $\Lambda<0$ or asymptotically AdS charged black hole spacetime,
equation (8) has two real positive roots $r_{h}$ and $r_{H}$,
where $r_{h}$ is the Cauchy horizon and $r_{H}$ is the event horizon \cite{ZZ3}.
From $r_{H}^2-2M r_{H}+Q^2=\Lambda r_{H}^4<0$, we have $r_{H}<M+\sqrt{M^2-Q^2}$.
And the mass ratio satisfies the upper bound
\begin{eqnarray}\label{AdSBH}
\frac{m^{+}_{hair}}{m^{-}_{hair}}=\frac{1}{\frac{r_{\gamma}}{r_{H}}-1}<\frac{1}{\frac{\frac{1}{2}(3M+\sqrt{9M^2-8Q^2})}{M+\sqrt{M^2-Q^2}}-1}
\leqslant2.
\end{eqnarray}

The upper bound corresponds to the case of $Q\rightarrow0$ and $\Lambda\rightarrow 0$.
The existence of this upper bound is natural since the
negative cosmological constant usually serves as a
potential to confine the matter field around the horizon.
Since we study the case of $0<r_{H}\leqslant r_{\gamma}<+\infty$, there is
$\frac{m^{+}_{hair}}{m^{-}_{hair}}=\frac{1}{\frac{r_{\gamma}}{r_{H}}-1}> 0$.
In all, we obtain bounds of the mass ratio
\begin{eqnarray}\label{AdSBH}
0<\frac{m^{+}_{hair}}{m^{-}_{hair}}< 2.
\end{eqnarray}

Now we show that this lower bound can be approached as $\Lambda\rightarrow -\infty$.
After choosing $Q\ll 1$, we solve $r^2-2Mr-\Lambda r^4=0$ to find the horizon $r_{H}$.
Since $-2Mr$ is the leading term in $r^2-2Mr-\Lambda r^4$ around $r\thickapprox 0$,
we have $r^2-2Mr-\Lambda r^4<0$ for r a little larger than 0.
There is also $r^2-2Mr-\Lambda r^4\rightarrow +\infty$ as $r\rightarrow +\infty$.
In the procedure of $\Lambda\rightarrow -\infty$, we divide the horizon
into three cases: $r_{H}\rightarrow 0$, $r_{H}\rightarrow \infty$ and $r_{H}\rightarrow C$,
where C is a non-zero constant.

In the cases of $r_{H}\rightarrow \infty$ and $\Lambda\rightarrow -\infty$, we have
\begin{eqnarray}\label{AdSBH}
r^2-2Mr\rightarrow +\infty
\end{eqnarray}
and
\begin{eqnarray}\label{AdSBH}
r^2-2Mr-\Lambda r^4\rightarrow +\infty
\end{eqnarray}
in contradiction with the equation $r^2-2Mr-\Lambda r^4=0$.

In another case of $r_{H}\rightarrow C\neq 0$ and $\Lambda\rightarrow -\infty$, we have
\begin{eqnarray}\label{AdSBH}
r^2-2Mr-\Lambda r^4\rightarrow C^2-2MC-\Lambda C^4\rightarrow+\infty
\end{eqnarray}
in contradiction with the equation $r^2-2Mr-\Lambda r^4=0$.

In a word, we have $r_{H}\rightarrow 0$ as $\Lambda\rightarrow -\infty$ and $Q\rightarrow 0$.
For the case of $Q\ll 1$ and M fixed, the lower bound can be approached as
\begin{eqnarray}\label{AdSBH}
\frac{m^{+}_{hair}}{m^{-}_{hair}}=\frac{1}{\frac{r_{\gamma}}{r_{H}}-1}
=\frac{1}{\frac{3M}{r_{H}}-1}\rightarrow 0~~~as~~~\Lambda\rightarrow -\infty.
\end{eqnarray}
Here the relation (19) shows that the linear hair of Maxwell field can invade the
Hod's lower bound $\frac{m^{+}_{hair}}{m^{-}_{hair}}\geqslant 1$. It implies that the Hod's bound may be invaded in the
asymptotically AdS Einstein-Born-Infeld hairy black holes
according to the fact that Born-Infeld field hair can be reduced to Maxwell field
in the linear limit. We will further check this in the following part.

\section{Hair mass bounds of Einstein-Born-Infeld black holes}

We should emphasize that the RN-(A)dS black hole is not hairy since the Maxwell field is
associated with a Gauss law. In this part, we extend the discussion to
Einstein-Born-Infeld hairy black holes with the Born-Infeld
factor associated with no conserved charge \cite{YSM,WYJ}.
The Lagrangian density for Born-Infeld theory is in the form
\begin{eqnarray}\label{AdSBH}
L_{BI}=\frac{1}{b^2}(1-\sqrt{1+\frac{b^2F^{\mu\nu}F_{\mu\nu}}{2}}).
\end{eqnarray}
Here b is the Born-Infeld factor parameter.
Mention that in the limit $b\rightarrow 0$,
this Lagrangian reduces to the Maxwell case
and properties of the RN black holes may also hold
in Born-Infeld hairy black holes at least for very small b.

Now we introduce the line element of Born-Infeld black holes with
non-zero cosmological constant $\Lambda$ as follows:
\begin{eqnarray}\label{AdSBH}
ds^{2}&=&-f_{EBI}(r)dt^{2}+f(r)_{EBI}^{-1}dr^{2}+r^{2}(d\theta^2+sin^{2}\theta d\phi^{2}),
\end{eqnarray}
where metric functions $f_{EBI}(r)=1-\frac{2M}{r}-\Lambda r^2+
\frac{2b^2r^2}{3}(1-\sqrt{1+\frac{Q^2}{b^2r^4}})+\frac{4Q^2}{3r^2}F[\frac{1}{4},\frac{1}{2},\frac{5}{4};-\frac{Q^2}{b^2r^4}]$,
where M is the ADM mass, Q is the charge and F is the hypergeometric function satisfying
$(\frac{1}{r}F[\frac{1}{4},\frac{1}{2},\frac{5}{4};-\frac{Q^2}{b^2r^4}])'_{r}=-\frac{1}{\sqrt{r^4+\frac{Q^2}{b^2}}}$ \cite{DG1,DG2,DG3}.
Around $x=0$, the hypergeometric function can be expanded as $F[a, b, c, x] =1+\frac{abx}{c}+\frac{a(1+a)b(1+b)x^2}{2c(1+c)}+o[x^3]$
(see Eq.15.7.1 of \cite{book}).
So we have $F[\frac{1}{4},\frac{1}{2},\frac{5}{4};-\frac{Q^2}{b^2r^4}]\rightarrow 1$
and $f_{EBI}(r)\rightarrow 1-\frac{2M}{r}+\frac{Q^2}{r^2}-\Lambda r^2$ as $r\rightarrow \infty$.
The photonsphere radius $r_{\gamma}$ is determined by the relation $2f_{EBI}(r_{\gamma})-r_{\gamma}f'_{EBI}(r_{\gamma})=0$ \cite{SH,YSM}
and the black hole event horizon $r_{H}$ can be numerically obtained from $f(r_{H})=0$ \cite{ZZ1,ZZ2,ZZ3}.

The energy density of the Born-Infeld hair is give by
$\rho(r)=-T_{t}^{t}=-\frac{(\frac{2b^2r^3}{3}(1-\sqrt{1+\frac{Q^2}{b^2r^4}})+\frac{4Q^2}{3r}F[\frac{1}{4},\frac{1}{2},\frac{5}{4};-\frac{Q^2}{b^2r^4}])'}{8\pi r^2}$.
The mass $m_{EBI}(r)$ of the Born-Infeld hair above the radius r is given by
\begin{eqnarray}\label{AdSBH}
m_{EBI}(r)=\int_{r}^{+\infty}4\pi r'^{2}\rho(r')dr'=\frac{1}{2}(\frac{2b^2r^3}{3}(1-\sqrt{1+\frac{Q^2}{b^2r^4}})+\frac{4Q^2}{3r}F[\frac{1}{4},\frac{1}{2},\frac{5}{4};-\frac{Q^2}{b^2r^4}]).
\end{eqnarray}

The hair mass ratio is
\begin{eqnarray}\label{AdSBH}
\frac{m^{+}_{EBI}}{m^{-}_{EBI}}=\frac{m_{EBI}(r_{\gamma})}{m_{EBI}(r_{H})-m_{EBI}(r_{\gamma})}
=\frac{1}{\frac{b^2r_{H}^3(1-\sqrt{1+\frac{Q^2}{b^2r_{H}^4}})+\frac{2Q^2}
{r_{H}}F[\frac{1}{4},\frac{1}{2},\frac{5}{4};-\frac{Q^2}{b^2r_{H}^4}]}{b^2r_{\gamma}^3(1-\sqrt{1+\frac{Q^2}{b^2r_{\gamma}^4}})+
\frac{2Q^2}{r_{\gamma}}F[\frac{1}{4},\frac{1}{2},\frac{5}{4};-\frac{Q^2}{b^2r_{\gamma}^4}]}-1}.
\end{eqnarray}

In the following, we calculate the hair mass ratio in the
Einstein-Born-Infeld genuine hairy black holes.
In the case of asymptotically flat space with
$\Lambda=0$, we find that the Hod's hair mass ratio
bound holds for various b and other parameters fixed.
For example, in the case $M=1.5$,~$Q=1$,~$\Lambda=0$ and various b
from 0.001 to 1000, we find that $\frac{m^{+}_{EBI}}{m^{-}_{EBI}}$
decreases as a function of b and the ratio
approaches the limit value 1.895 for large b.
We show part of the numerical data in Table I and
it can be easily seen from the table that
the mass ratio is above the Hod's lower bound.
Since the mass ratio with $b=2$ almost reaches the lowest value, we fix
$b=2$ to check the Hod's bound in the following.
\renewcommand\arraystretch{1.7}
\begin{table} [h]
\centering
\caption{The hair mass ratio $\frac{m^{+}_{EBI}}{m^{-}_{EBI}}$ with $M=1.5$,~$Q=1$,~$\Lambda=0$ and various b}
\label{address}
\begin{tabular}{|>{}c|>{}c|>{}c|>{}c|>{}c|>{}c|>{}c|>{}c|>{}c|>{}c|}
\hline
$~b~$ &~0.1~& ~0.3~& ~0.5~& ~0.7~& ~0.9~& ~1.0~& ~2.0~& ~4.0~& ~6.0~\\
\hline
$~\frac{m^{+}_{EBI}}{m^{-}_{EBI}}~$ & ~2.288~ & ~1.952~& ~1.916~& ~1.905~& ~1.901~& ~1.890~& ~1.895~& ~1.895~& ~1.895~\\
\hline
\end{tabular}
\end{table}

We also find that the Hod's bound holds with different
values of M. In Table II, we show that the ratio is above
the Hod's lower bound in the case of $Q=1$,~$b=2$,~$\Lambda=0$ and
various $M$. According to Table II, the ratio increases
as a function of M and the smallest ratio
$\frac{m^{+}_{EBI}}{m^{-}_{EBI}}\thickapprox 1.218$ is obtained
in the case of $M\thickapprox1.0$, which corresponds
to the extremal black hole solution with $M=Q$.
\renewcommand\arraystretch{1.7}
\begin{table} [h]
\centering
\caption{The hair mass ratio $\frac{m^{+}_{EBI}}{m^{-}_{EBI}}$ with $Q=1$,~$b=2$,~$\Lambda=0$ and various $M$}
\label{address}
\begin{tabular}{|>{}c|>{}c|>{}c|>{}c|>{}c|>{}c|>{}c|>{}c|>{}c|>{}c|>{}c|>{}c|}
\hline
$~M~$ &~1.0~& ~1.1~& ~1.2~& ~1.3~& ~1.4~& ~1.5~& ~1.6~& ~1.7~& ~1.8~& ~1.9~& ~2.0~\\
\hline
$~\frac{m^{+}_{EBI}}{m^{-}_{EBI}}~$ & ~1.218~ & ~1.668~& ~1.780~& ~1.837~& ~1.872~& ~1.895~& ~1.913~& ~1.926~& ~1.936~& ~1.934~& ~1.950~\\
\hline
\end{tabular}
\end{table}

We also show cases of $M=1.5$,~$b=2$,~$\Lambda=0$ and various $Q$ in Table III.
Here the ratio decreases with respect to the charge Q and in the case of
the extremal black hole solution with $Q\thickapprox1.5$,
the smallest ratio is $\frac{m^{+}_{EBI}}{m^{-}_{EBI}}\thickapprox 1.147$ above the Hod's lower bound.
We further numerically check for the
parameters in a larger range and find that
the Hod's bound $\frac{m^{+}_{hair}}{m^{-}_{hair}}\geqslant 1$ \cite{SH} holds in
the asymptotically flat Einstein-Born-Infeld hairy black hole in accordance with results in \cite{YSM}.
\renewcommand\arraystretch{1.7}
\begin{table} [h]
\centering
\caption{The hair mass ratio $\frac{m^{+}_{EBI}}{m^{-}_{EBI}}$ with $M=1.5$,~$b=2$,~$\Lambda=0$ and various $Q$}
\label{address}
\begin{tabular}{|>{}c|>{}c|>{}c|>{}c|>{}c|>{}c|>{}c|>{}c|>{}c|}
\hline
$~Q~$ &~0.1~& ~0.3~& ~0.5~& ~0.7~& ~0.9~& ~1.1~& ~1.3~& ~1.5~\\
\hline
$~\frac{m^{+}_{EBI}}{m^{-}_{EBI}}~$ & ~1.999~ & ~1.993~& ~1.980~& ~1.958~& ~1.922~& ~1.861~& ~1.736~& ~1.147~\\
\hline
\end{tabular}
\end{table}

Now we will further show that the Hod's bound also holds in the asymptotically dS
Einstein-Born-Infeld hairy black holes.
The data in Table IV represents the mass ratio $\frac{m^{+}_{EBI}}{m^{-}_{EBI}}$
with respect to the positive cosmological constants $\Lambda$
in the case of $M=1.5$,~$Q=1$ and $b=2$. We see that
the mass ratio increases as we choose a larger cosmological constant
with other parameters fixed and
the smallest ratio can be obtained in the case
of $\Lambda=0$ or flat space.
Since the Hod's lower bound holds in the asymptotically
flat Einstein-Born-Infeld hairy black holes, we can conclude
that the Hod's lower bound also holds in the
asymptotically dS Einstein-Born-Infeld hairy black holes.
\renewcommand\arraystretch{1.7}
\begin{table} [h]
\centering
\caption{The hair mass ratio $\frac{m^{+}_{EBI}}{m^{-}_{EBI}}$ with $M=1.5$,~$Q=1$,~ $b=2$ and various positive $\Lambda$}
\label{address}
\begin{tabular}{|>{}c|>{}c|>{}c|>{}c|>{}c|>{}c|>{}c|>{}c|>{}c|>{}c|>{}c|}
\hline
$~\Lambda~$ &~0~& ~0.001~& ~0.005~& ~0.010~& ~0.015~& ~0.016~& ~0.017~& ~0.018~& ~0.019~& ~0.020~\\
\hline
$~\frac{m^{+}_{EBI}}{m^{-}_{EBI}}~$ & ~1.895~ & ~1.942~& ~2.169~& ~2.633~& ~3.724~& ~4.190~& ~4.907~& ~6.241~& ~10.422~& ~117.500~\\
\hline
\end{tabular}
\end{table}

However, we find that the Hod's lower bound $\frac{m^{+}_{hair}}{m^{-}_{hair}}\geqslant 1$ 
can be invaded in the AdS background.
With $M=1.5$,~$Q=1$,~$b=2$ and different values of $\Lambda$, we find
that the hair mass ratio is below the Hod's lower bound
for very negative cosmological constants as can be seen in Table V.
\renewcommand\arraystretch{1.7}
\begin{table} [h]
\centering
\caption{The hair mass ratio $\frac{m^{+}_{EBI}}{m^{-}_{EBI}}$ with $M=1.5$,~$Q=1$,~$b=2$ and various negative $\Lambda$}
\label{address}
\begin{tabular}{|>{}c|>{}c|>{}c|>{}c|>{}c|>{}c|>{}c|}
\hline
$~\Lambda~$ &~0~& ~-0.02~& ~-0.04~& ~-0.06~& ~-0.08~& ~-0.10~\\
\hline
$~\frac{m^{+}_{EBI}}{m^{-}_{EBI}}~$ & ~1.895~ & ~1.381~& ~1.153~& ~1.015~& ~0.921~& ~0.851~\\
\hline
\end{tabular}
\end{table}

We have further checked for the whole 4 dimensional
parameter space in a very large range and the properties
are qualitatively the same as cases in Tables (I,~II,~III,~IV,~V).
In summary, we show that the Hod's bound holds in
Einstein-Born-Infeld hairy black holes with non-negative cosmological constants.
In contrast, the Hod's bound can be invaded in the
asymptotically AdS Einstein-Born-Infeld hairy black holes.
Our results imply that the more negative cosmological
constant makes the Born-Infeld hair more easier to
condense in the near horizon region.
Moreover, we conjecture that the Hod's bound
may be also invaded in other AdS hairy black holes.
For AdS black holes, a potential well in the near horizon
region forms due to the AdS boundary \cite{sn1},
which provides the confinement of the scalar field
and may make the scalar hair easier to condense in the
near horizon well \cite{sn2}. And another possible method to confine
the scalar hair in the near horizon region is enclosing
the black hole in a scalar reflecting box \cite{box1,box2,box3,box4,box5}.
We also mention that Skyrme hairs with
cosmological constants have been studied \cite{SC1,SC2,SC3}.
We plan to examine effects of cosmological constants on
scalar hair and Skyrme hair distributions in the next work.
Moreover, there is no scalar hair theorem in regular neutral reflecting stars \cite{Hod1,SBS}
and static scalar fields can condense around charged reflecting stars
\cite{Hod2,Hod3,Hod4,EMP,YP1,YP2,YP3,YP4,YP5}.
So it is also very interesting to extend the discussion to the reflecting star background.

\section{Conclusions}

We studied mass distribution of linear hair in RN black holes and genuine hair in
Einstein-Born-Infeld theory with various cosmological constants.
We used the event horizon and the photonsphere to divide the hair into two parts and
obtained lower bounds for the mass ratio.
We found that the Hod's lower bound obtained in asymptotically
flat gravity also holds in the asymptotically dS Einstein-Born-Infeld hairy black holes.
In contrast, the Hod's lower bound can be invade in the
asymptotically AdS Einstein-Born-Infeld hairy black holes.
Our results showed that the more negative cosmological constants make the
Born-Infeld hair easier to condense in the near horizon area.
And we further conjectured that effects of cosmological constants on hair distribution may be
qualitatively the same in other hairy black holes.

\begin{acknowledgments}

We would like to thank the anonymous referees for the constructive suggestions to improve the manuscript.
This work was supported by the Shandong Provincial Natural Science Foundation of China under Grant No. ZR2018QA008.

\end{acknowledgments}


\begin{thebibliography}{99}



\bibitem{RR}
R. Ruffini and J. A. Wheeler,Introducing the black hole, Phys. Today 24, 30(1971).



\bibitem{JDB}
J. D. Bekenstein, Black holes and entropy, Phys. Rev. D 7, 2333(1973).


\bibitem{PW}
Panofsky, W.K.H,Needs Versus Means In High-energy Physics,
Phys. Today 33, 24(1980).





\bibitem{WI1}
W. Israel,Event horizons in static vacuum space-times, Phys. Rev. 164, 1776(1967).





\bibitem{BC}
B. Carter,Axisymmetric Black Hole Has Only Two Degrees of Freedom, Phys. Rev. Lett. 26, 331(1971).





\bibitem{SW}
S. W. Hawking,Black holes in general relativity, Commun. Math. Phys. 25, 152(1972).








\bibitem{DC2}
D. C. Robinson,Uniqueness of the Kerr black hole, Phys. Rev. Lett. 34, 905(1975).



\bibitem{JI}
J. Isper,Electromagnetic Test Fields Around a Kerr-Metric Black Hole,
Phys. Rev. Lett. 27, 529 (1971).










\bibitem{JE1}
J. E. Chase, Commun. Math. Phys. 19, 276(1970).

\bibitem{JE2}
J. D.Bekenstein,Transcendence of the law of baryon-number conservation in black hole physics, Phys. Rev. Lett. 28, 452(1972).



\bibitem{JE3}
C. Teitelboim,Nonmeasurability of the baryon number of a black-hole, Lett. Nuovo Cimento 3, 326(1972).














\bibitem{BM2}
M. Heusler,A No hair theorem for selfgravitating nonlinear sigma models, J. Math. Phys. 33, 3497(1992).


\bibitem{BM3}
D.Sudarsky,A Simple proof of a no hair theorem in Einstein Higgs theory, Class. Quantum Grav. 12, 579(1995).


















\bibitem{JH1}
J. Hartle,Long-range neutrino forces exerted by kerr black holes, Phys. Rev. D 3, 2938(1971).



\bibitem{JH2}
C. Teitelboim,Nonmeasurability of the lepton number of a black hole, Lett. Nuovo Cimento 3, 397(1972).








\bibitem{PBH1}
P. Bizo$\acute{n}$,Colored black holes, Phys. Rev. Lett 64, 2844(1990).



\bibitem{PBH2}
M. S. Volkov and D. V. Gal'tsov, Sov. J. Nucl. Phys. 51, 1171(1990).



\bibitem{PBH3}
H. P. Kuenzle and A. K. M. Masood-ul-Alam,Spherically symmetric static SU(2) Einstein Yang-Mills fields,
J. Math. Phys. 31,928(1990).















\bibitem{H1}
G. Lavrelashvili and D. Maison,Regular and black hole solutions of Einstein Yang-Mills Dilaton theory, Nucl. Phys. B 410, 407
(1993).




\bibitem{H2}
P. Bizo$\acute{n}$ and T. Chamj,Gravitating skyrmions, Phys. Lett B 297, 55(1992).



\bibitem{H3}
Serge Droz, Markus Heusler, Norbert Straumann,New black hole solutions with hair, Phys. Lett. B 268,
371(1991).




\bibitem{H4}
M.Heusler, S. Droz, and N. Straumann,Stability analysis of selfgravitating skyrmions, Phys. Lett. B 271, 61(1991).







\bibitem{H6}
B. R. Greene, S. D. Mathur, and C. O'Neill,Eluding the no hair conjecture:
Black holes in spontaneously broken gauge theories, Phys. Rev.
D 47, 2242(1993).



\bibitem{H7}
T. Torii, K. Maeda, and T. Tachizawa,NonAbelian black holes and catastrophe theory. 1. Neutral type, Phys. Rev. D 51,1510(1995).






\bibitem{H8}
N. Straumann and Z.-H Zhou,Instability of a colored black hole solution, Phys. Lett. B 243, 33
(1990).




\bibitem{H9}
N. E. Mavromatos and E. Winstanley,	Aspects of hairy black holes in spontaneously broken Einstein
Yang-Mills systems: Stability analysis and entropy considerations, Phys. Rev. D 53,
3190(1996).




\bibitem{H10}
M. S. Volkov and D. V. Gal'tsov,Gravitating nonAbelian solitons and black holes with Yang-Mills fields,
Phys. Rept. 319, 1(1999).




\bibitem{H12}
G. V. Lavrelashvili and D. Maison,A Remark on the instability of the Bartnik-McKinnon solutions, Phys. Lett. B 343,
214(1995).



\bibitem{H13}
Yves Brihaye, Carlos Herdeiro, Eugen Radu, D. H. Tchrakian,
Skyrmions, Skyrme stars and black holes with Skyrme hair in five spacetime dimension, JHEP11(2017)037.






\bibitem{H15}
C. A. R. Herdeiro and E. Radu,Kerr black holes with scalar hair, Phys. Rev. Lett. 112, 221101 (2014).





\bibitem{H17}
P. V. P. Cunha, C. A. R. Herdeiro, E. Radu, and H. F. R¡äunarsson, Shadows of Kerr black holes with scalar hair,Phys.
Rev. Lett. 115, 211102(2015).


\bibitem{H18}
Y. Brihaye, C. Herdeiro, and E. Radu,Inside black holes with synchronized hair, Phys. Lett. B 760, 279(2016).






\bibitem{H19}
J. D. Bekenstein,Black hole hair:twenty--five years after, arXiv:gr-qc/9605059.


\bibitem{H20}
Carlos A. R. Herdeiro, Eugen Radu,Asymptotically flat black holes with scalar hair: A review,
Int.J.Mod.Phys.D 24(2015)09,1542014.
















\bibitem{DNH}
D. N$\acute{u}\tilde{n}$ez, H. Quevedo, and D. Sudarsky,Black Holes Have No Short Hair, Phys. Rev. Lett.
76, 571(1996).


\bibitem{SHA}
Shahar Hod,A no-short scalar hair theorem for rotating Kerr black holes, Class.Quant.Grav. 33(2016)114001.



\bibitem{SH}
S. Hod,Hairy Black Holes and Null Circular Geodesics, Phys. Rev. D 84, 124030 (2011).


\bibitem{YSM}
Yun Soo Myung, Taeyoon Moon,Hairy mass bound in the Einstein-Born-Infeld black hole,
Phys. Rev. D 86.084047.











\bibitem{ZZ1}
Vitor Cardoso, Madalena Lemos, Miguel Marques,
On the instability of Reissner-Nordstr$\ddot{o}$m black holes in de Sitter backgrounds,
Phys. Rev. D 80(2009)127502.



\bibitem{ZZ2}
Zhiying Zhu, Shao-Jun Zhang, C.E. Pellicer, Bin Wang, Elcio Abdalla,
Stability of Reissner-Nordstr$\ddot{o}$m black hole in de Sitter background under charged scalar perturbation,
Phys. Rev. D 90(2014)044042.


\bibitem{ZZ3}
Bin Wang, Chi-Yong Lin, C. Molina,
Quasinormal behavior of massless scalar field perturbation in Reissner-Nordstr$\ddot{o}$m anti-de Sitter spacetimes,
Phys. Rev. D 70(2004)064025.




\bibitem{EW}
Elizabeth Winstanley,Classical Yang-Mills black hole hair in anti-de Sitter space,
Lect.Notes Phys.769:49-87,2009.









\bibitem{WYJ}
Weiping Yao,Jiliang Jing,Holographic entanglement entropy in insulator/superconductor transition with Born-Infeld electrodynamics,
JHEP 05(2014)058.


\bibitem{DG1}
T. K. Dey,Born-Infeld black holes in the presence of a
cosmological constant, Phys. Lett. B 595(2004)484.



\bibitem{DG2}
S. Fernando,Thermodynamics of Born-Infeld-anti-de Sitter
black holes in the grand canonical ensemble, Phys. Rev. D 74(2006)104032.



\bibitem{DG3}
Y. S. Myung, Y. W. Kim and Y. J. Park,Thermodynamics and phase transitions in the
Born-Infeld-anti-de Sitter black holes, Phys. Rev. D
78(2008)084002.







\bibitem{book}
M. Abramowitz and I. A. Stegun, Handbook of Mathematical Functions (Dover Publications, New York, 1970).














\bibitem{sn1}
Yunqi Liu, De-Cheng Zou, Bin Wang, Signature of the Van der Waals like small-large charged AdS black hole
phase transition in quasinormal modes, JHEP 09(2014)179.



\bibitem{sn2}
Pablo Bosch, Stephen R. Green, and Luis Lehner, Nonlinear Evolution and Final Fate of Charged AntiCde Sitter
Black Hole Superradiant Instability, Phys. Rev. Lett. 116(2016)141102.






\bibitem{box1}
Pallab Basu, Chethan Krishnan, P. N. Bala Subramanian,Hairy Black Holes in a Box,
JHEP 1611(2016)041.

\bibitem{box2}
Nicolas Sanchis-Gual,Juan Carlos Degollado,Pedro J.Montero,Jos A. Font,Carlos Herdeiro,Explosion and final
state of an unstable Reissner-Nordstr$\ddot{o}$m black hole, Phys. Rev. Lett. 116(2016)141101.

\bibitem{box3}
Sam R Dolan,Supakchai Ponglertsakul,Elizabeth Winstanley, Stability of black holes in Einstein-charged scalar
field theory in a cavity, Phys. Rev. D 92(2015)124047.



\bibitem{box4}
Yan Peng, Bin Wang, Yunqi Liu ,
On the thermodynamics of the black hole and hairy black hole transitions in the asymptotically flat spacetime with a box,
Eur.Phys.J. C 78(2018)176.



\bibitem{box5}
Yan Peng,Studies of a general flat space/boson star transition model in a box through a language similar to holographic superconductors,
JHEP 07(2017)042.





\bibitem{SC1}
Noriko Shiiki, Nobuyuki Sawado,Black hole skyrmions with negative cosmological constant,
Phys. Rev. D 71(2005)104031.

\bibitem{SC2}
Ilya Perapechka,Yakov Shnir,Generalized Skyrmions and hairy black holes in asymptotically $AdS_{4}$ spacetime,
Phys. Rev. D 95(2017)025024.

\bibitem{SC3}
Marco Astorino, Fabrizio Canfora, Marcela Lagos, Aldo Vera,
Black hole and BTZ-black string in the Einstein-SU(2) Skyrme model,
Phys. Rev. D 97(2018)124032.





\bibitem{Hod1}
S. Hod,No-bomb theorem for charged Reissner-Nordstr$\ddot{o}$m black holes, Physics Letters B 718, 1489 (2013).

\bibitem{SBS}
Srijit Bhattacharjee, Sudipta Sarkar, No-hair theorems for a static and stationary reflecting star, Phys. Rev.D 95(2017)084027.


\bibitem{Hod2}
S.Hod, Charged massive scalar field configurations supported by a spherically symmetric
charged reflecting shell, Physics Letters B 763, 275 (2016).



\bibitem{Hod3}
S.Hod, Marginally bound resonances of charged massive scalar fields in the background of
a charged reflecting shell, Physics Letters B 768(2017)97-102.


\bibitem{Hod4}
Shahar Hod,Charged reflecting stars supporting charged massive scalar field configurations,
European Physical Journal C 78, 173 (2017).


\bibitem{EMP}
Elisa Maggio,Paolo Pani,Valeria Ferrari,Exotic Compact Objects and How to Quench their Ergoregion Instability,
Phys. Rev. D 96(2017)104047.


\bibitem{YP1}
Yan Peng, Scalar field configurations supported by charged compact reflecting stars in a curved spacetime,
Physics Letters B 780(2018)144-148.



\bibitem{YP2}
Yan Peng, Studies of scalar field configurations supported by reflecting shells in the AdS spacetime,
Eur. Phys. J. C 78(2018)680.



	
\bibitem{YP3}
Yan Peng,Static scalar field condensation in regular asymptotically AdS reflecting star backgrounds, Physics Letters B 782(2018)717-722.

	
\bibitem{YP4}
Yan Peng,Scalar condensation behaviors around regular Neuman reflecting stars, Nuclear Physics B 934(2018)459-465.



\bibitem{YP5}
Yan Peng, On instabilities of scalar hairy regular compact reflecting stars, arXiv:1810.04102.



\end{thebibliography}
\end{document}